\documentclass[aps,prb,twocolumn,superscriptaddress,groupedaddress,floatfix,longbibliography]{revtex4-2}

\usepackage{amsmath,amssymb} 
\usepackage{physics}
\usepackage{bm} 
\usepackage{graphicx} 
\usepackage{comment} 
\usepackage{textcomp} 

\usepackage{enumitem}
\setlist{noitemsep,leftmargin=*,topsep=0pt,parsep=0pt}

\usepackage{xcolor} 
\definecolor{lightgray}{gray}{0.6}
\definecolor{medgray}{gray}{0.4}
\usepackage{tikz}
\usepackage{comment}
\usepackage{hyperref}
\hypersetup{
colorlinks=true,
urlcolor= blue,
citecolor=blue,
linkcolor= blue,
}

\begin{document}

\title{Valence Bond Glass and Glassy Spin Liquid in Disordered Frustrated Magnets}


\author{Soumyaranjan Dash}
\affiliation{Department of Physical Sciences, Indian Institute of Science Education and Research (IISER) Mohali, Sector 81, S.A.S. Nagar, Manauli PO 140306, India}

\author{Vansh Narang}
\affiliation{Department of Physical Sciences, Indian Institute of Science Education and Research (IISER) Mohali, Sector 81, S.A.S. Nagar, Manauli PO 140306, India}

\author{Sanjeev Kumar}
\email{sanjeev@iisermohali.ac.in}
\affiliation{Department of Physical Sciences, Indian Institute of Science Education and Research (IISER) Mohali, Sector 81, S.A.S. Nagar, Manauli PO 140306, India}

\begin{abstract}

The absence of conventional magnetic order together with anomalous low-temperature magnetic heat capacity is often interpreted as evidence for quantum spin liquid ground states in frustrated magnets. Using a recently developed semiclassical Monte Carlo approach, we show that similar thermodynamic signatures arise in the highly frustrated regime of the disordered spin-$1/ 2$  $J_1–J_2$ Heisenberg model on the square lattice. By analyzing the freezing parameters, the distribution of spin–spin correlations, and the specific heat, we identify the ground state as a valence-bond glass that melts into a glassy spin liquid at finite temperatures. 
We show that the low-temperature specific-heat anomaly originates from collective singlet excitations, and consequently it is insensitive to external magnetic fields. This leads to a robust experimental signature of the valence bond glass phase and a completely new interpretation of the thermodynamic data on disordered spin-liquid candidate materials.

\end{abstract}

\date{\today}
\maketitle

{\it Introduction:--}
Quantum spin liquids (QSLs) are highly sought-after quantum phases of matter predicted to exist in frustrated magnetic systems \cite{Anderson1973, Savary2017, Zhou2017, Broholm2020, Balents2010, Wen2019}. Beyond their fundamental significance as exotic phases of quantum matter characterized by long-range entanglement and fractionalized excitations, QSLs hold promise for applications in quantum information and computation technologies \cite{Balents2010, Wen2002, Semeghini2021, Kitaev2003, Basov2017}. The absence of conventional order in QSLs makes their experimental identification intrinsically challenging. In practice, the presence of a QSL phase is inferred indirectly from characteristic features in thermodynamic observables that are attributed to fractionalized excitations \cite{Wang2025, Knolle2015, Nasu2017, Jin2023, Li2020, Theveniaut2017, Eschmann2020}. Consequently, the possibility of alternative mechanisms producing similar thermodynamic signatures calls such inferences into question.

One proposed route to realizing a QSL phase is through the introduction of disorder in frustrated magnetic materials \cite{Motrunich2005, Watanabe2014, Kawamura2019, Uemetsu2018}. This mechanism can be understood in terms of disorder enhancing the effectiveness of magnetic frustration in suppressing the conventional magnetic order, thereby driving the system towards a QSL ground state. Experimental evidence of such disorder-induced behavior, which is referred to as QSL-like, has been reported in various frustrated spin-$1/2$ magnets, such as, Sr$_2$Cu(Te$_{1-x}$W$_x$)O$_6$ \cite{Yoon2021}, YbMgGaO$_4$ \cite{Zhu2017, Kimchi2018}, Li$_4$CuTeO$_6$ \cite{Khatua2022}, 1T-TaS$_{2-x}$Se$_x$ \cite{Murayama2020}. From a theoretical perspective, the thermodynamic properties of the relevant interacting disordered models remain largely inaccessible due to the lack of methods that treat strong correlations and quenched disorder on equal footing \cite{Byczuk2009, Redka2024, Prozorov2024}. Therefore, characterizing and understanding the nature of disorder-induced quantum phases in models for frustrated magnets is an outstanding theoretical problem.

In this work, we study the effect of quenched disorder on the spin-$1/2$ $J_1-J_2$ Heisenberg model on a square lattice using a recently developed semiclassical approach \cite{Dash2026}. We establish valence-bond glass (VBG) as the ground state of the disordered model near the maximal frustration point. The VBG is characterized by the freezing of spin-spin correlations and the anomalous low-temperature specific heat originating from collective singlet excitations \cite{Klanjvsek2017, Carlo2011}.
The temperature dependence of the freezing parameters and specific heat helps us identify a thermal crossover from the VBG state to a glassy spin liquid (GSL) phase and finally to a conventional paramagnet (PM).
The qualitative validity of the semiclassical approximation is tested by comparing with the ED results on small clusters. The presence of non-spinon excitations gives rise to experimentally testable signatures of the VBG phase, and allows for a completely new interpretation of the experimental data on disordered frustrated magnets.

{\it Model and methods:--}
We consider the disordered variant of the spin-$1/2$ $J_1$-$J_2$ Heisenberg model on a square lattice. The Hamiltonian is given by,
\begin{equation}
\begin{aligned}
H =\;
& \sum_{\langle ij \rangle} J_{1, ij}\, ~ {\bf s}_i \cdot {\bf s}_j
+ \sum_{\langle \langle ij \rangle \rangle} J_{2,ij}\, ~ {\bf s}_i \cdot {\bf s}_j, \\
\end{aligned}
\label{eq:Ham1}
\end{equation}

\noindent 
where ${\bf s}_i$ represents a spin-$1/2$ operator, and $J_{1, ij}$, $J_{2,ij}$ are the values of the nearest neighbor (nn) and next nn (nnn) exchange parameters, selected from box distributions of widths $2J_1\Delta$ and $2 J_2 \Delta$ centered at $J_1$ and $J_2$, respectively. The single (double) angular bracket indicates the summation over nn (nnn) pairs of sites. The main focus of this study is to understand the nature of quantum phases induced by the quenched disorder at the maximum frustration point $J_2/J_1 = 0.5$. All energies are measured in unit of $J_1$, which is set to unity.

Applying the recently developed semiclassical (SC) approach to the Hamitonian Eq. (\ref{eq:Ham1}), we arrive at the following model of Ising spins and link variables \cite{Dash2026}: 
\begin{equation}
\begin{aligned}
H_{\rm sc} =\;
& \sum_{\langle ij \rangle \in \rm{U}} J_{1,ij} ~ S_i S_j
+ \sum_{\langle \langle ij \rangle \rangle \in \rm{U}} J_{2,ij} ~ S_i S_j \\
& + \sum_{\langle ij \rangle \in \rm{C}} E_{1,ij} ~ b_{ij}
+ \sum_{\langle \langle ij \rangle \rangle \in \rm{C}} E_{2, ij} ~ b_{ij}.
\end{aligned}
\label{eq:Ham2}
\end{equation}
\noindent
In the above, $S_i$ are Ising variables with values $\pm 1/2$ and $b_{ij}$ are bond variables taking values $0$ or $1$. The energies associated with bond variables, for $n=1,2$, are given by \cite{Dash2026},

\begin{equation}
E_{n,ij} =
-\frac{3 J_{n,ij}}{4}
\left[
 \frac{1-e^{-J_{n,ij}/T}}{1 + 3e^{-J_{n,ij}/T}}
\right].
\label{eq:Eb}
\end{equation}

\newcommand{\hcylpair}{%
\tikz[baseline=0.5ex, scale=0.35, line width=0.5pt]{
  \draw (0,0) ellipse (0.12 and 0.04);
  \draw (0,1.0) ellipse (0.12 and 0.04);
  \draw (-0.12,0) -- (-0.12,1.0);
  \draw (0.12,0) -- (0.12,1.0);

  \draw (0.5,0) ellipse (0.12 and 0.04);
  \draw (0.5,1.0) ellipse (0.12 and 0.04);
  \draw (0.38,0) -- (0.38,1.0);
  \draw (0.62,0) -- (0.62,1.0);
}}

The Hamiltonian Eq. (\ref{eq:Ham2}) is simulated using the standard Markov chain Monte Carlo process. The Monte Carlo updates incorporate several local moves to ensure efficient exploration of configuration space. These include single spin-flip updates, and the creation and annihilation of singlets on nn and nnn links \cite{Dash2026}. In addition, proposals of rotations of nn singlet pairs from horizontal to vertical and vice versa, which may be represented as $\hcylpair \leftrightarrow \raisebox{-0.3ex}{\rotatebox{90}{\hcylpair}}$, facilitate efficient sampling of resonating singlet configurations. We also retain the quantum trait of singlets by imposing the local constraint that any given lattice site participates in the formation of at most one dimer. At each temperature step, the system is equilibrated through $N_{{\rm eq}}\sim 10^5$ Monte Carlo Sweeps (MCS) before computing the averages on $N_{{\rm av}}\sim 10^5$ MCS. In order to ensure that the results do not depend on a specific realization of the exchange coefficients, we perform configuration averages over $\sim 100$ independent realizations. The results obtained within the semiclassical approach are also compared against those obtained by exact diagonalization (ED) on $N$-site clusters with $N=16$. The Hamiltonian matrix is constructed in the $\{ \hat{\textbf{s}}_{i , z}\}$ basis, and the ground state is obtained by diagonalizing the block corresponding to $\hat{\textbf{S}}_{{\rm total}} = 0$. 

{\it Results and Discussion:--}
We begin by discussing the influence of disorder on the distributions of the nn spin-spin correlations defined as, 
\begin{eqnarray}
\langle {\bf s}_i \cdot {\bf s}_j \rangle & = &
\dfrac{1}{N_{{\rm av}}} \sum_{k = 1}^{N_{{\rm av}}} ({\bf s}_i \cdot {\bf s}_j)_k, \nonumber \\ 
({\bf s}_i \cdot {\bf s}_j)_k & = & \begin{cases}
S_i S_j, & \text{if} \ \ i,j \in \text{U} \\
E_{1,ij}/J_{1,ij} & \text{if } i,j \in \text{C} \\
0 & \text{otherwise}.
\end{cases}
\label{eq:Corr}
\end{eqnarray}

\begin{figure}
\includegraphics[width=0.98 \columnwidth,angle=0,clip=true]{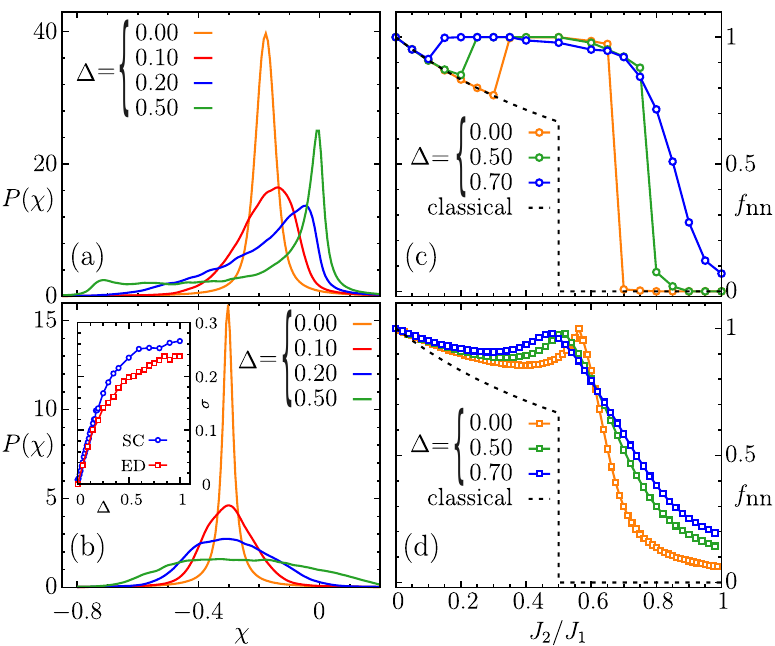}
 \caption{Distributions of nn correlation, $P(\chi)$, for different values of disorder strength $\Delta$ obtained from, (a) the MC simulations on the SC model at $T = 0.10$, (b) the ED ground state on $4 \times 4$ lattice. Inset in (b) shows the $\Delta$-dependence of the variance, $\sigma$, of the distributions shown in (a) and (b).  Variation of the relative contribution of the nn component of energy with $J_2/J_1$ for different $\Delta$ obtained from, (c) the SC simulations and (d) the ED ground state. Dashed lines in (c) and (d) is the result for $\Delta = 0$ for classical spins which does not display any upturn at intermediate values of $J_2/J_1$.}
\label{sfig1}   
\end{figure}

\noindent  In the context of the semiclassical approach, $\langle ... \rangle$ denotes the averaging over Monte Carlo configurations and $({\bf s}_i\cdot {\bf s}_j)_k$ the nn spin-spin correlations in the $k^{th}$ configuration. We compute the distribution function of the nn correlations, $P(\chi) = \sum_{\langle ij \rangle} \delta(\chi - \langle {\bf s}_i \cdot {\bf s}_j \rangle)$, by using the standard Lorentzian approximation for the Dirac-delta functions with broadening $0.02$.  Spin-spin correlations for each pair of spins are bound to be in the interval $[-0.75, 0.25]$. In the clean limit, all nn correlations lead to identical values due to translational invariance, as reflected in the sharp peak in the distribution function (see Fig. \ref{sfig1}(a)). The distribution broadens with increasing $\Delta$, indicating the loss of translational invariance. The broadening can be understood as the preference of singlets to remain frozen on links with stronger $J_1$. The appearance of peaks near $-0.75$ and zero is reflective of the semiclassical assumption of perfect singlets on the nn links.

To assess the validity of the SC approximation in capturing the effects of quenched disorder, we compute nn spin–spin correlations and their distributions using ED. The ED results on the $ 4 \times 4$ lattice exhibit a similar disorder-induced broadening of the distribution [Fig. \ref{sfig1}(b)]. The $\Delta $ dependence of the width of the distribution obtained at low temperature from the SC simulations shows good agreement with that obtained from the ED (inset of Fig. \ref{sfig1}(b)). Our ED results for the distributions agree well with previous studies \cite{Uematsu2018}.  We further test the validity of our SC approach by computing the fractional contribution of nn energy defined as, $f_{nn} = |E_{nn}|/(|E_{nn}| + |E_{nnn}|),$ where $E_{nn}$ and $E_{nnn}$ refer to the average energy from nn and nnn terms in the SC Hamiltonian. In a purely classical approximation, $f_{nn}$ shows a monotonic decrease with increasing $J_2$, before vanishing at $J_2 = 0.5$ (dashed lines in Figs. \ref{sfig1}(c) and \ref{sfig1}(d)). The SC approach shows a non-monotonic variation with larger values in the intermediate $J_2/J_1$ regime. This qualitative deviation from the classical result is verified by computing $f_{nn}$ in the exact ground states obtained by ED. The ED results are consistent with a non-monotonic variation of $f_{nn}$, with the peak near $J_2/J_1 = 0.6$ (Fig. \ref{sfig1}(d)). In both the SC and the ED results, the intermediate higher-$f_{nn}$ region broadens with increasing $\Delta$. For $f_{nn}$, the ED results for $N=16$ compare well with the numbers reported for $N=40$ \cite{Richter2010}. This qualitative agreement between ED and the SC approach further supports the validity of the latter in the presence of disorder. A key advantage of the SC method is its ability to access system sizes nearly two orders of magnitude larger than those accessible via ED, enabling highly reliable extrapolation to the thermodynamic limit.

\begin{figure}
\includegraphics[width=0.98 \columnwidth,angle=0,clip=true]{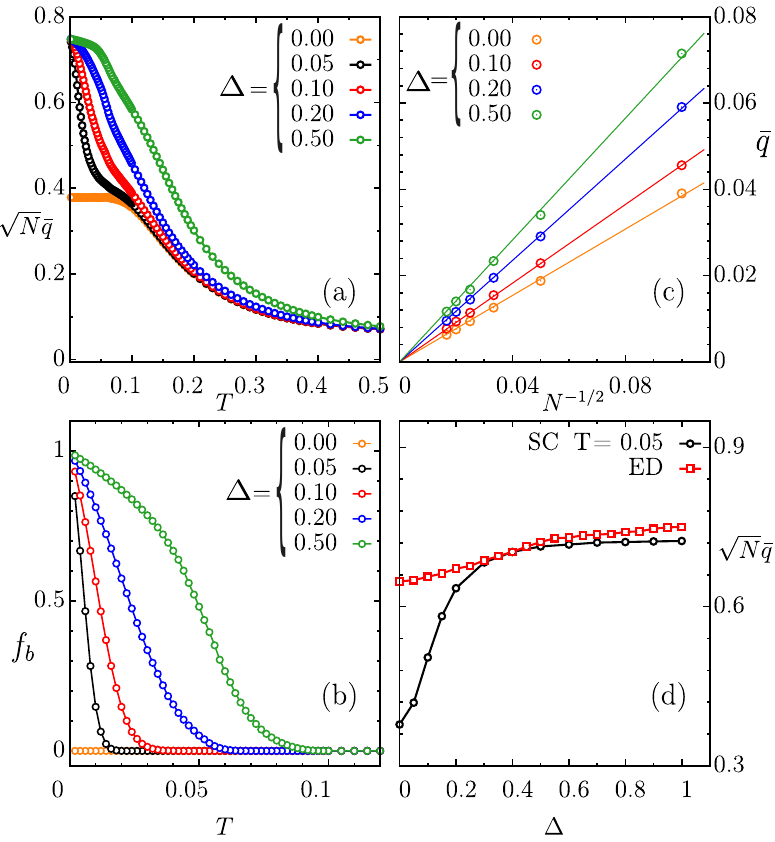}
 \caption{Temperature dependence of, (a) the freezing parameter, defined in Eq. (\ref{eq:qbar}), and (b) Fraction of completely frozen NN singlets, defined in Eq. (\ref{eq:fb}), for different values of $\Delta$ obtained via SC simulations. (c) The scaling of $\bar q$ with $\sqrt{N}$ for different $\Delta$ from the SC simulations. (d) Comparison of the $\Delta$-dependence of $\sqrt{N}~ \bar q$, defined in Eq. (\ref{eq:qbar}), between the SC simulations and the ED on $N=16$.}
\label{sfig2}   
\end{figure}

For a quantitative analysis of non-magnetic quantum phases induced by disorder, we utilize SC simulations to compute the freezing parameter, $\bar q$, and the fraction of fully frozen singlets, $f_{b}$, defined as,
\begin{eqnarray}
\bar q(T) & = & \frac{1}{N}
\sqrt{ \sum_{i\neq j}
\left\langle \mathbf{s}_i\cdot\mathbf{s}_j \right\rangle^{2} }, 
\label{eq:qbar}
\end{eqnarray}

\begin{eqnarray}
f_b(T) & = & \frac{1}{N}
 \sum_{\langle ij \rangle}
\prod_{k=1}^{N_{{\rm av}}}
b_{ij}^{(k)}.
\label{eq:fb}
\end{eqnarray}

\noindent 
In Eq. (\ref{eq:qbar}) above, the sum is over all $(i,j)$ pairs with $i \neq j$. The sum in Eq. (\ref{eq:fb}) is only over nn sites and the product is over the $N_{{\rm av}}$ MC steps where $b_{ij}^{(k)}$ denotes the value of the variable $b_{ij}$ in configuration $k$. Note that in any phase where only short-range correlations survive, $\bar q$ should scale as $1/\sqrt{N}$ since the number of terms contributing to the sum is proportional to $N$.  In contrast, in a long-range ordered phase, $\bar q$ should be constant as the number of contributing terms is $O(N^2)$. Furthermore, since all correlations vanish at high temperatures, $\bar q$ is expected to go to zero at large $T$. Therefore, $\sqrt{N} ~ \bar q$ serves as an indicator for a liquid-like phase with only short-range correlations. 

The temperature dependence of $\sqrt{N} ~ \bar q$ for different $\Delta$ is shown in Fig. \ref{sfig2}(a). Note that the value of $\bar q$ is larger in a phase where the dimers remain frozen on specific nn links, compared to a phase having dimer locations fluctuating across MC configurations. At $T=0$, the numbers in the two cases are easy to estimate: $\sqrt{N} ~ \bar q \approx 0.75$ for frozen singlets and $\sqrt{N} ~ \bar q \approx 0.375$ when the singlets fluctuate between nn bonds. At $\Delta = 0$, the $T$-dependence of $\sqrt{N}~ \bar q$ shows a crossover from a QSL ground state to a PM phase (Fig. \ref{sfig2}(a)). For finite values of $\Delta$, $\sqrt{N}~ \bar q$ displays a two-step reduction with increasing $T$. This indicates a two-step release of the entropy with increasing temperature. The higher value at low $T$ signifies a phase with singlet correlations frozen on selected links. Such a phase with no local magnetic moments and frozen spin-spin correlations is termed VBG \cite{Niestemski2009, devries2010, Singh2010, Watanabe2018, Riedl2019, Lee2021, Coomer2013, Mustonen2022, Mclaughlin2010}. The interpretation of the phase with larger $\bar q$ as VBG is further confirmed by the survival of the fraction of frozen singlets $f_b$ at low $T$ and finite $\Delta$. The disappearance of $f_b$ at finite $T$, which also correlates well with the location of the first inflection point in $\sqrt{N}~ \bar q$, indicates the melting of the VBG phase (Fig. \ref{sfig2}(b)). However, the values of $\sqrt{N} ~ \bar q$ in this melted phase are still higher than those in the SL phase at equivalent temperatures. Consequently, the intermediate $T$ regime is best described as a glassy spin liquid (GSL) \cite{Tarzia2008, devries2010}. The thermal evolution at finite disorder can be summarized as VBG to GSL to PM.

As noted above, $\bar q$ for any short-range ordered phase should scale as $1/\sqrt{N}$. We find that the expected scaling holds for all values of $\Delta$ at low temperature (Fig. \ref{sfig2}(c)). The slope of the linear scaling is the value of $\sqrt{N} ~ \bar q$ plotted in Fig. \ref{sfig2}(a). Ideally, one would like to confirm this scaling from the ED data. However, the range of lattice sizes accessible within ED do not lead to a reliable scaling behavior \cite{Richter2010, Uematsu2018}. Nevertheless, we can directly compare the disorder dependence of $\sqrt{N}~ \bar q$ obtained from the ED calculations on $4 \times 4$ lattice with that from the SC simulations. A similar increase is noted in both results (see Fig. \ref{sfig2}(d)). The values differ for small $\Delta$ due to the presence of stronger correlations in the ground states obtained through ED. This difference is related to the fact that ED solutions retain quantum-correlations better as reflected in the difference between the peak locations of the distributions for $\Delta = 0$ (Figs. \ref{sfig1}(a) and \ref{sfig1}(b)). However, for larger values of disorder $\bar q$ is dictated by the stronger correlations on specific links. Therefore, the accuracy of the SC approach improves rapidly with increasing $\Delta$. 

\begin{figure}
\includegraphics[width=0.98 \columnwidth,angle=0,clip=true]{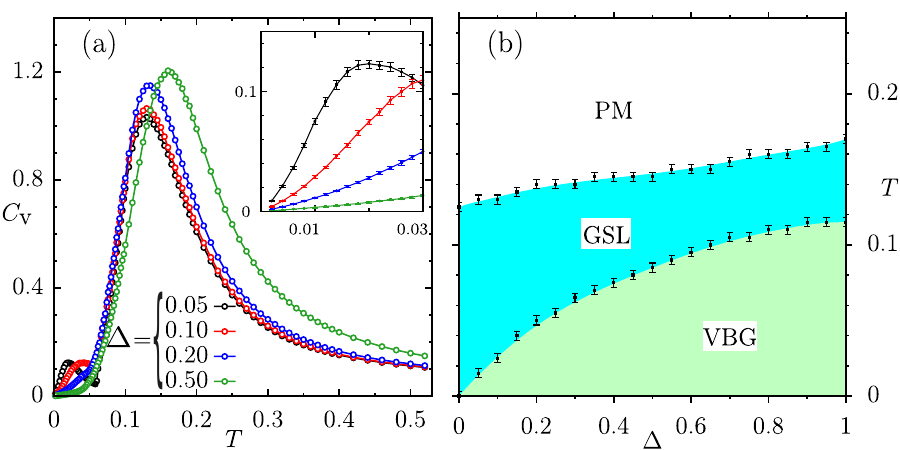}
 \caption{(a) Temperature dependence of specific heat for different values of $\Delta$ obtained from SC simulations. Inset in (a) shows the nearly linear behavior at low $T$. (b) Phase diagram in the $T$-$\Delta$ plane displaying the crossover lines between VBG, GSL and PM.  The boundary between VBG and GSL is inferred from $f_b(T)$ and that between GSL and PM tracks the location of the higher-$T$ peak in the specific heat.}
\label{sfig3}   
\end{figure}

In order to identify thermodynamic signatures of the VBG state, we track the $T$-dependence of specific heat, $C_V = d\langle E \rangle/dT $, for different $\Delta$ obtained via SC simulations (Fig. \ref{sfig3}(a)). For $\Delta = 0$, a broad peak characteristic of the crossover from the QSL to PM phase is obtained. Note that neither SC simulations nor ED on small clusters are capable of capturing spinon excitations, and therefore no low-$T$ feature is present in the clean limit. However, $C_V$ displays an algebraic rise at low-$T$ for small non-zero $\Delta$, up to a $T$ scale proportional to $\Delta$. The slope at low $T$ decreases with increasing $\Delta$, and the specific-heat for large $\Delta$ displays the conventional disordered-exponential behavior.

Within the SC approach, low-energy excitations that induce the algebraic $T$-dependence in the specific heat at finite $\Delta$ are identified as local singlet-pair rotations, represented  as $\hcylpair \leftrightarrow \raisebox{-0.3ex}{\rotatebox{90}{\hcylpair}}$. Note that at $\Delta = 0$, such pair-rotations in a random singlet state do not cost any energy and therefore do not contribute to $C_V$. For small nonzero $\Delta$, such MC moves cost low energy and are accepted in the Metropolis algorithm at finite $T$. This leads to an increase in the average energy, and hence the specific heat, without breaking any singlets. For larger values of $\Delta$, the singlet excitations are strongly suppressed at low $T$, justifying a much weaker rise of specific heat.  The presence of the algebraic rise of $C_V$ with $T$ suggests that the relevance of the collective singlet excitations, which are believed to exist along with the spinon excitations in the clean model \cite{Kotov1999}, is enhanced due to disorder. This is consistent with the interpretation that spinons are Anderson-localized in the presence of disorder and their contribution to the low-$T$ specific heat is strongly suppressed \cite{Pan2022, Kim2022, Lavarelo2013}.

The fact that low energy excitations in the VBG phase do not involve breaking of singlets has important experimental implications. Being within the $S_{{\rm total}} = 0$ part of the Hilbert space, such collective singlet excitations do not couple to external magnetic fields. Consequently, the low-$T$ specific heat should remain independent of magnetic field. This is in sharp contrast to the field response of specific heat arising from spinons. In a disorder-free QSL system, the collective singlet excitations are subdominant compared to the spinon excitations. Therefore, the low-$T$ specific heat is affected by external magnetic fields \cite{Li2015, Liu2022}. However, in the presence of disorder the spinons undergo Anderson localization and the collective singlet excitations become relevant. This signature has already been reported in disordered double-perovskite magnets, Sr$_2$Cu(Te$_{1-x}$W$_x$)O$_6$, where the coefficient of linear specific heat did not change in the presence of magnetic fields of strength up to $9~$T \cite{Watanabe2018}. Interestingly, complete field independence of specific heat has also been reported in hyper-kagome, Na$_4$Ir$_3$O$_8$,  which are known to be intrinsically disordered \cite{Okamoto2007, Singh2013}. 
Our work shows that such field-independence of low-$T$ specific heat is a robust signature of the VBG phase. This general interpretation is further supported by the NMR data on Na$_4$Ir$_3$O$_8$ displaying a perfect correlations between slow dynamics and algebraic $T$ dependence of specific heat \cite{Shockley2015}.


The temperature and disorder dependence of $\sqrt{N} ~ \bar q$, the fraction of frozen singlets, $f_b$, and the specific heat is used to characterize the phases as VBG, GSL and PM as summarized in the $T-\Delta$ phase diagram (see Fig. \ref{sfig3}(b)). We find that the ground state in the presence of disorder is best described as VBG, with a characteristic melting temperature that increases monotonically with $\Delta$. The melted VBG state is labeled as GSL as it has characteristics that are intermediate between VBG and QSL. Note that a QSL ground state does not exist in the presence of quenched disorder. However, a variant of QSL with slower dynamics -- termed as GSL-- exists over a wide regime in the $T$-$\Delta$ plane. 

{\it Conclusions:--}
Using a recently developed SC approximation, we show that the ground state of the disordered square-lattice $J_1-J_2$ Heisenberg model near $J_2/J_1 = 0.5$ is VBG. The VBG state, characterized by the absence of magnetic moments and frozen singlet correlations, melts into a GSL phase upon increasing temperature, and finally into a conventional PM. The use of the SC approach in the presence of quenched disorder is justified by an explicit comparison with the ED calculations. In particular, we find that the quantitative accuracy of the SC approach increases in the presence of disorder. The access to large lattice sizes allows us to explicitly track the thermodynamic behavior of the model. For temperature scales below the disorder strength, we find a nearly linear $T$-dependence of specific-heat that originates from excitations related to singlet-pair rotations in the random singlet state. Unlike the spinon contribution, the collective singlet excitations are not sensitive to weak external magnetic fields. Therefore, the VBG state induced by the quenched disorder can be identified in experiments from the field-independence of low-$T$ specific heat. The discovery of a thermodynamic indicator of VBG state is a result with broad implications for existing and future experimental data on QSL candidate materials. The GSL phase reported here may be easier to realize in
disordered frustrated magnets as compared to the conventional QSL state. Moreover, the inherent slow dynamics of the GSL is likely to suppress quantum decoherence, implying improved functionalities for quantum computations. Our results call for a reassessment of experimental claims of QSL behavior in disordered frustrated magnets and position GSLs as a distinct and highly relevant class of quantum-disordered matter.

{\it Acknowledgements:--}
We thank Yogesh Singh for many useful discussions. We acknowledge the use of the HPC facility at IISER Mohali. S. D. and V. N. acknowledge IISER Mohali for support through the institute fellowship.

\bibliography{library_rand_j1j2}



\end{document}